\documentclass[aps,prx,amsmath,superscriptaddress,twocolumn,bibnotes,longbibliography,floatfix]{revtex4-2}
\usepackage{amsmath}
\usepackage{amsfonts}
\usepackage{graphicx}
\usepackage{color}
\usepackage{mathtools}
\usepackage{bm}
\usepackage[colorlinks=true,linkcolor=blue,urlcolor=blue,citecolor=blue]{hyperref}
\usepackage{orcidlink}
\DeclareMathOperator{\sgn}{sgn}
\newcommand{\vex}[1]{\bm{\mathrm{#1}}}
\newcommand{\ket}[1]{\left| {#1} \right\rangle}
\newcommand{\bra}[1]{\left\langle {#1} \right|}

\begin{document}

	\title{Double quantum dots as dynamical impurities in one-dimensional transport}

	\author{Yang-Zhi~Chou~\orcidlink{0000-0001-7955-0918}}\email{yzchou@umd.edu}
	\affiliation{Condensed Matter Theory Center and Joint Quantum Institute, Department of Physics, University of Maryland, College Park, Maryland 20742, USA}

	\date{\today}

	\begin{abstract}
		We study single-electron scattering in a tight-binding chain capacitively coupled to double quantum dots, which act as dynamical two-level-system impurities with electrically tunable level splittings. For a single impurity, we solve the scattering problem exactly and show Feshbach-like perfect elastic reflection, whose resonant condition distinguishes the initial two-level state. For several impurities in a strongly reflecting chain, transmission peaks at an intermediate mean level splitting. At small splittings, energy exchange with the impurities can weaken constructive interference between backscattered paths, consistent with the initial rise in transmission. As excitation channels close, virtual transitions can enhance reflection and drive transmission below its purely elastic value $T_0$. At larger splittings, these virtual processes weaken and transmission approaches $T_0$. We discuss how gate-controlled transmission measurements in quantum-dot arrays could test these predictions.
	\end{abstract}

	\maketitle

\textit{Introduction.---} Anderson localization arises from quantum interference between repeated elastic scatterings and suppresses diffusion in disordered systems \cite{AndersonPW1958a}. At finite temperatures, intrinsic electron--electron interactions can induce inelastic scattering that dephases time-reversed electronic paths and suppresses weak localization \cite{AltshulerBL1982}. Interactions, however, don't necessarily destroy localization, as exemplified by many-body localization in isolated disordered systems \cite{GornyiIV2005,BaskoDM2006,NandkishoreR2015,AbaninDA2019}.

An external quantum environment supplies additional dynamical degrees of freedom with which electrons can exchange energy, allowing inelastic scattering and dephasing. Counterintuitively, coupling to these degrees of freedom can also favor localization. Several bosonization studies have shown that dissipative environments (e.g., coupling to a heat bath) can destabilize a Luttinger liquid and produce signatures reminiscent of disorder-induced localization \cite{CazalillaMA2006,MajumdarS2023,MajumdarS2023a,Bouverot-DupuisO2026a}. Additionally, dissipation can suppress quantum tunneling, as demonstrated by Caldeira and Leggett, and suppressing electron tunnelings \cite{SchmidA1983,CaldeiraAO1981,RistivojevicZ2008}. The distinct effects, such as dephasing and enhancing reflection, due to inelastic scattering motivate several basic questions: What are the conditions that inelastic scattering may enhance/induce localization? What is the minimal setting that demonstrates these effects?

Motivated by the questions raised above, in this Letter, we study electron scattering in a tight-binding chain capacitively coupled to double quantum dots (DQDs), as a minimal model for studying dynamical impurities in one-dimensional (1D) transport. Each DQD acts as a dynamical two-level system (TLS) impurity with an electrically tunable level splitting. For a single impurity, we solve the single-electron scattering problem exactly and find that coupling to a bound state in a closed channel can produce perfect elastic reflection, reminiscent of the Feshbach resonance. For several spatially separated DQDs, our multichannel calculations show that transmission through a strongly reflecting chain varies \textit{nonmonotonically} with the mean level splitting. The initial rise in transmission is consistent with energy exchange weakening constructive interference between backscattered paths. As the splitting increases further, excitation channels close and transmission can fall below its value for the same chain with the dynamical coupling switched off.
We propose gate-controlled transmission measurements in quantum-dot arrays to test these predictions and discuss their implications for other types of dynamical impurities, such as local oscillators.

	\begin{figure}[t]
		\includegraphics[width=0.4\textwidth]{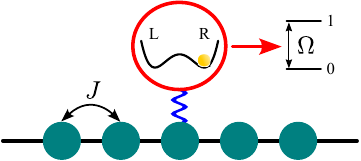}
		\caption{Single dynamical impurity Setup. We consider a 1D tight-binding channel (teal dots) capacitively coupled to a DQD (red dot). The hopping strength of the 1D tight-binding model is denoted by $J$, and the electron interacts with the DQD at the impurity site. The DQD can be viewed as a double well confining potential with one electron, and it can be viewed as a local TLS. Concomitantly, the interaction between 1D electrons and DQD can be transformed into Eq.~\eqref{Eq:H_I_TLS}, including a transverse $\sigma^x$ coupling as discussed in the main text.
		}
		\label{Fig:Setup}
	\end{figure}

	\textit{Model. ---} We consider a 1D tight-binding model capacitively coupled to DQDs \cite{SchomerusH2002,ElzermanJM2003,HuY2007,KwapinskiT2012,HellM2014,PettaJR2004,GormanJ2005,PeterssonKD2010,HaldarS2025}. The single impurity setup is illustrated in Fig.~\ref{Fig:Setup}. The 1D tight-binding model is described by
	\begin{align}
		\label{Eq:H_e}\hat{H}_e=&-J\sum_{j}\left(c^{\dagger}_{j+1}c_j+\text{H.c.}\right),
	\end{align}
	where $c_j$ is the electron annihilation operator at site $j$, and $J>0$ denotes the hopping strength.
	A DQD is made of a double-well confining potential and an electron that can hop between the right and left wells. The DQD can be modeled as follows:
	\begin{align}
		\label{Eq:H_DQD}\hat{H}_{\rm DQD}=&\sum_{\alpha}\left[\begin{array}{c}
			\frac{\epsilon_{\alpha}}{2}\left(d^{\dagger}_{R,\alpha}d_{R,\alpha}-d^{\dagger}_{L,\alpha}d_{L,\alpha}\right)\\[2mm]
			+\omega_{\alpha}\left(d^{\dagger}_{L,\alpha}d_{R,\alpha}+\text{H.c.}\right)
		\end{array}\right],
	\end{align}
	where $\alpha$ denotes the index of the DQD, $d_{R,\alpha}$ ($d_{L,\alpha}$) denotes the electron annihilation operator in the right (left) minimum of the $\alpha$th DQD, $\epsilon_{\alpha}$ is the detuning, $\omega_{\alpha}$ is the hybridization between the right and the left wells in DQD. We require only one electron in each DQD, i.e., $d^{\dagger}_{R,\alpha}d_{R,\alpha}+d^{\dagger}_{L,\alpha}d_{L,\alpha}=1$. Finally, the electrons in the 1D tight-binding model and the DQDs interact via Coulomb force, which can be modeled by \cite{KwapinskiT2012,HellM2014}
	\begin{align}
		\label{Eq:H_I}\hat{H}_{I,\text{DQD}}\!=&\!\sum_{\alpha}\!c^{\dagger}_{j_{\alpha}}c_{j_{\alpha}}\!\!\left(V_{R,\alpha}d^{\dagger}_{R,\alpha}d_{R,\alpha}\!+\!V_{L,\alpha}d^{\dagger}_{L,\alpha}d_{L,\alpha}\right).
	\end{align}
	In the above expression, $V_{R,\alpha}$ and $V_{L,\alpha}$ couple the charge at site $j_\alpha$ to the charges in the right and left wells of the $\alpha$th DQD, respectively.

	These DQDs can be described by local TLS impurities. Equation~(\ref{Eq:H_DQD}) can be expressed by
	\begin{align}
		\label{Eq:H_TLS}\hat{H}_{\rm DQD}\rightarrow\hat{H}_{\rm TLS}=&\sum_{\alpha}\frac{\Omega_{\alpha}}{2}\left(\hat{1}_{\alpha}+\hat{\sigma}^{z}_{\alpha}\right),
	\end{align}
	where the energy splitting $\Omega_{\alpha}=2\sqrt{\omega^2_{\alpha}+\epsilon^2_{\alpha}/4}$, $\hat{\sigma}^{\mu}_{\alpha}$ is the $\mu$-component Pauli matrix of the $\alpha$th impurity, and $\hat{1}_{\alpha}$ denotes the 2-by-2 identity matrix. For each TLS impurity, the ground state and the excited states are described by the quantum number $m_{\alpha}=0$ and $m_{\alpha}=1$, respectively. The total TLS energy is given by $\sum_{\alpha}m_{\alpha}\Omega_{\alpha}$.

	Using the same TLS basis, the coupling Hamiltonian [Eq.~(\ref{Eq:H_I})] becomes
	\begin{align}
		\label{Eq:H_I_TLS}\hat{H}_{I,\text{DQD}}\rightarrow\hat{H}_{I}=&\sum_{\alpha}\!c^{\dagger}_{j_{\alpha}}c_{j_{\alpha}}\!\left(U_{\alpha}\hat{1}_{\alpha}+g_{\alpha}\hat{\sigma}^x_{\alpha}+g'_{\alpha}\hat{\sigma}_{\alpha}^z\right),
	\end{align}
	where
	\begin{subequations}\label{Eq:Int_para}
		\begin{align}
			U_{\alpha}=&\frac{V_{R,\alpha}+V_{L,\alpha}}{2},\\
			g_{\alpha}=&-\frac{(V_{R,\alpha}-V_{L,\alpha})\omega_{\alpha}}{\Omega_{\alpha}},\\
			g'_{\alpha}=&\frac{(V_{R,\alpha}-V_{L,\alpha})\epsilon_{\alpha}}{2\Omega_{\alpha}}.
		\end{align}
	\end{subequations}
	The $U_\alpha$ and $g'_\alpha$ terms describe channel-symmetric and channel-antisymmetric elastic scattering, respectively. The transverse coupling $g_\alpha$ \cite{AnokhinAO2006,RammalH2026,BashanN2024a,TulipmanE2024a} flips the TLS and vanishes when the interdot hybridization $\omega_\alpha$ is zero. Additionally, we consider local external potentials on the impurity sites (e.g., by plunger gates). We absorb any local gate potentials at the impurity sites into $U_\alpha$, so the total on-site potential can have either sign. The full Hamiltonian is $\hat{H}_e+\hat{H}_{\rm TLS}+\hat{H}_{I}$ [Eqs.~\eqref{Eq:H_e}, \eqref{Eq:H_TLS}, and \eqref{Eq:H_I_TLS}].

	\textit{Single-electron single-impurity scattering.--- }We first discuss single electron scattering off a single TLS impurity. For simplicity, we assume the single impurity is located at site $j=0$ and drop the impurity index. In Eq.~(\ref{Eq:H_I_TLS}) with only one impurity, the scattering potential is given by $\begin{bmatrix}
		u_1 & g\\
		g & u_0
	\end{bmatrix}$, where $u_1=U+g'$ and $u_0=U-g'$. The total energy is given by $E=\mathcal{E}_m+m\Omega$, where $m$ indicates the TLS quantum number, and $\mathcal{E}_m$ is the electron energy. Besides extended scattering states, the single-impurity Hamiltonian can support discrete electron--TLS bound states localized near the impurity. Here we consider scattering states with an electron incident from $j=-\infty$; the discrete bound states do not constitute additional outgoing channels. We solve this problem exactly using the Lippmann–Schwinger equation. See Appendix~\ref{App:Single_TLS} for a detailed derivation. We summarize the results in the following.

	First, we consider $-2J<\mathcal{E}_1<\mathcal{E}_0<2J$, corresponding to plane wave solutions in both electron scattering channels (i.e., both channels are open). We can define velocities $v_0=\sqrt{4J^2-\mathcal{E}_0^2}$ and $v_1=\sqrt{4J^2-\mathcal{E}_1^2}$. The incident wave carries a TLS quantum number $s$, and we define $\bar{s}=1-s$ for the other TLS quantum number. The transmission ($T$) and reflection (R) probabilities are given by
	\begin{subequations}
		\begin{align}
			T_{s\rightarrow s}=&\frac{1+\delta_{\bar{s}}^2}{(1+\gamma-\delta_0\delta_1)^2+(\delta_0+\delta_1)^2},\\
			R_{s\rightarrow s}=&\frac{(\gamma-\delta_0\delta_1)^2+\delta_s^2}{(1+\gamma-\delta_0\delta_1)^2+(\delta_0+\delta_1)^2},\\
			T_{s\rightarrow \bar s}=&R_{s\rightarrow \bar s}=\frac{\gamma}{(1+\gamma-\delta_0\delta_1)^2+(\delta_0+\delta_1)^2},
		\end{align}
	\end{subequations}
	where $\gamma=g^2/(v_0v_1)$, $\delta_0=u_0/v_0$, and $\delta_1=u_1/v_1$. Scattering is elastic when the TLS remains in $s$ and inelastic when it flips to $\bar s$. For $U=g'=0$, elastic transmission dominates for $\gamma\ll1$, elastic reflection dominates for $\gamma\gg1$, and the TLS-flip probability is maximal at $\gamma=1$.

	\begin{figure}[t]
		\includegraphics[width=0.3\textwidth]{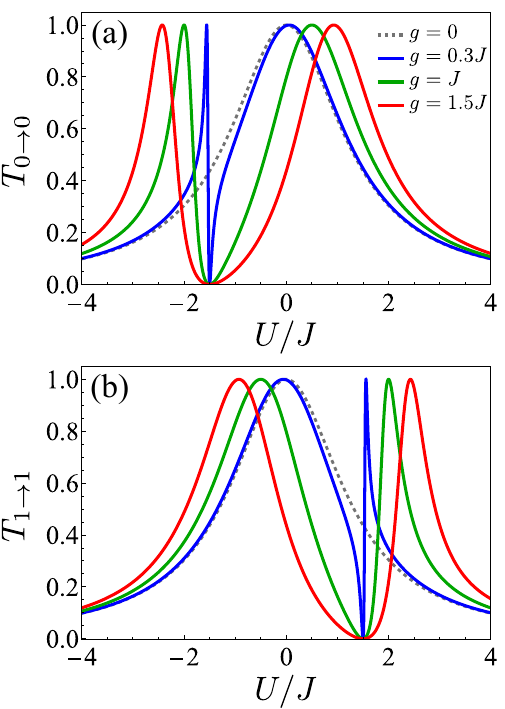}
		\caption{Single-impurity transmission probability with only one open channel. Several representative values of $g$ are chosen. (a) $s=0$ and $\mathcal{E}_0=-1.5J$. For $g\neq 0$, perfect reflection ($T_{0\rightarrow 0}=0$) takes place at $U=-1.5J$. (b) $s=1$ and $\mathcal{E}_1=1.5J$. For $g\neq 0$, perfect reflection ($T_{1\rightarrow 1}=0$) takes place at $U=1.5J$. $\Omega=J$ and $g'=0$ for all curves
		}
		\label{Fig:RT_one_close}
	\end{figure}

	Now, we discuss the case that only the incident channel is open. When $\Omega$ is sufficiently large, the non-incident electron channel becomes closed, forming an exponentially decaying evanescent state. In this situation, the TLS impurity scattering can be viewed as an effective elastic scattering with transmission and reflection probabilities given by
	\begin{align}\label{Eq:R_T_one_open}
		T_{s\rightarrow s}=\frac{v_s^2}{v_s^2+\mathcal{U}_{\text{eff},s}^2},R_{s\rightarrow s}=\frac{\mathcal{U}_{\text{eff},s}^2}{v_s^2+\mathcal{U}_{\text{eff},s}^2},
	\end{align}
	where
	\begin{align}\label{Eq:U_eff}
		\mathcal{U}_{\text{eff},s}=&u_s+\frac{g^2}{\sgn(\mathcal{E}_{\bar{s}})\sqrt{\mathcal{E}_{\bar{s}}^2-4J^2}-u_{\bar{s}}}.
	\end{align}
	Equation~(\ref{Eq:U_eff}) is the effective elastic scattering potential of channel $s$, incorporating both the bare elastic scattering $u_s$ and the virtual process to the closed channel $\bar{s}$. (See also Ref.~\cite{BandopadhyayS2008} for the continuum case.) We assume $g\neq 0$ for the following discussions. When  $\sgn(\mathcal{E}_{\bar{s}})\sqrt{\mathcal{E}_{\bar{s}}^2-4J^2}=u_{\bar{s}}$, the $\mathcal{U}_{\text{eff},s}$ diverges, leading to a perfect reflection, i.e., $R_{s\rightarrow s}=1$. Meanwhile, $\mathcal{U}_{\text{eff},s}=0$ corresponds to a perfect transmission $T_{s\rightarrow s}=1$. Particularly, the resonant condition of the perfect reflection depends mainly on the initial TLS state: For $s=0$ ($s=1$), perfect reflection resonant condition requires $u_1=U+g'<0$ ($u_0=U-g'>0$). The resonant reflection is reminiscent of the Feshbach resonance: the transverse coupling $g$ couples the propagating channel to an impurity-bound state of the uncoupled closed channel.

	In Fig.~\ref{Fig:RT_one_close}, we plot the transmission probability as a function of $U$ for several representative values of $g$, assuming $g'=0$ (i.e., $u_0=u_1=U$) for simplicity. In the absence of dynamical coupling ($g=0$), the elastic scattering produces a Lorentzian profile with perfect transmission ($T_{s\rightarrow s}=1$) at $U=0$ and $1/U^2$ decay for large $|U|$. For a nonzero $g$, the perfect transmission peak splits into two, and a resonant perfect reflection ($T_{s\rightarrow s}=0$) emerges at a finite value of $U$, whose sign depends on the incident TLS quantum number $s$, as discussed in the previous paragraph. Moreover, the value of $g$ crucially controls the width of the resonant reflection. With a finite $g'$, the qualitative results remain, but the locations of the transmission peaks and dips may vary. Since the value of $U$ can be tuned by the local electric gate (e.g., in quantum dot array setup), the resonant perfect reflection can in principle read out the initial TLS state $s$.

	\textit{Single-electron multi-impurity scattering. ---} In the presence of multiple impurities, the single-electron scattering becomes highly complex, as the number of channels is $2^N$ for an $N$-impurity problem. Each channel is marked by a binary vector $\vex{m}=(m_1,m_2,\dots,m_N)$. The single-electron problem here can be mapped to a coupled multi-channel problem exactly. The key idea is that the $g_{\alpha}$ term in Eq.~\eqref{Eq:H_I_TLS} can be viewed as a hopping between two channels with $m_{\alpha}=0$ and $m_{\alpha}=1$.

	For a single-electron problem with quenched onsite disorder, Anderson localization is inevitable in the infinite length limit. The general expectation is that dynamical impurities, such as the TLS inelastic scatterings here, can weaken localization. In finite-size systems with finite impurities, localization may not be sharply established, so we focus on how dynamical effects influence the transmission and reflection in the following discussions.

	For an electron incident from the left on three impurities at $x_1<x_2<x_3$, consider two third-order terms in the Born series for the reflection amplitude. The paths $P=(2,1,3)$ and $Q=(3,1,2)$ visit the impurities in opposite orders and interfere constructively for elastic scattering~\cite{GornyiIV2007}. If the TLS at $x_2$ flips, the flip occurs first along $P$ and last along $Q$. Both paths reach the same final channel, but the change in electron wave vector from $k$ to $k'$ gives them a relative phase $\Delta\varphi=2(k-k')(x_3-x_1)$. Energy exchange can therefore weaken their constructive interference. At larger TLS splittings, inelastic channels close, and virtual scattering through a closed channel can enhance elastic reflection, as in the single-impurity problem. The above arguments motivate a nonmonotonic transport dependence as the TLS splittings are tuned, which we investigate numerically.

	\begin{figure}[t]
		\includegraphics[width=0.3\textwidth]{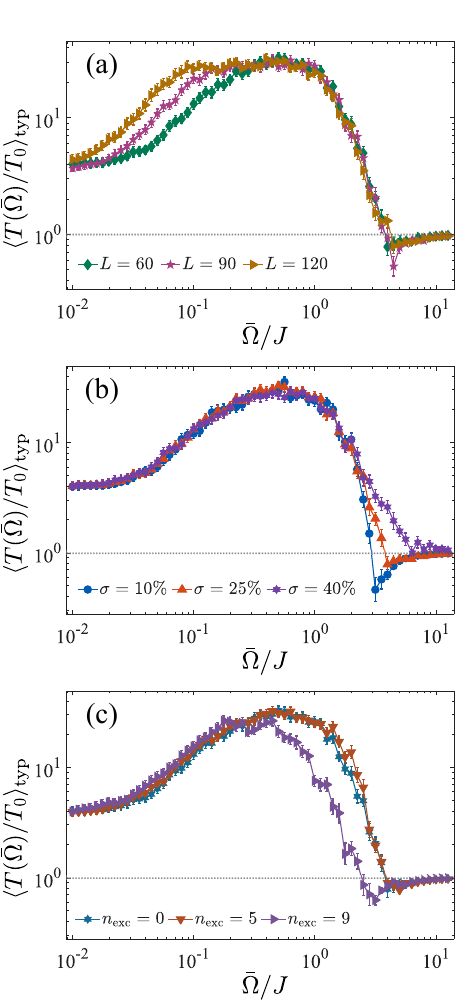}
		\caption{Rescaled transmission $T(\bar\Omega)/T_0$ as a function of mean TLS splitting $\bar\Omega$. We consider a 9-impurity system with incident electron energy $\mathcal E_{\vex s}=0.5J$. The total chain length is $L$. Each curve is the geometric mean of the per-realization ratio over 100 realizations. Error bars show one standard error on either side of the mean of $\ln[T(\bar\Omega)/T_0]$, with both limits transformed back to the plotted scale. The reference $T_0$ is calculated with $g_\alpha=g'_\alpha=0$ at the same impurity positions, elastic potentials, and incident electron energy. Every realization satisfies $0<T_0\le10^{-3}$ before dynamical couplings are added. The dotted horizontal line marks $T/T_0=1$. (a) Dependence of chain length $L$. We consider $\sigma=0.25$ and TLS ground state as the incident channel for the three cases here. (b) Dependence of $\sigma$.  We choose the same elastic potential profile and set $L=60$ for all three cases. The incident channel is the TLS ground state. (c) Dependence of incident channel. We choose $\sigma=0.25$ and $L=60$ for all three cases. $n_{\rm exc}$ denotes the number of excited TLS impurities.
		}
		\label{Fig:T_Omega}
	\end{figure}

	We solve the multichannel scattering problem using a sparse transfer-matrix formulation~\cite{KhomyakovPA2005,GrothCW2014} that retains both open and closed channels (Appendix~\ref{App:Transfer_matrix}). Since our goal is to examine how localization is impacted by the dynamical impurities, we select elastic potentials such that every realization has a \textit{small} TLS-free transmission $T_0(\mathcal E_{\vex s})\le10^{-3}$, calculated with $g_\alpha=g'_\alpha=0$ at the actual incident electron energy and the same elastic configuration. Then, we add TLS $\sigma^x$ couplings and the TLS splittings randomly. Particularly, $\Omega_{\alpha}= \bar\Omega(1+\sigma z_{\alpha})$, where the random variable $z_{\alpha}$ has zero mean and unit root mean square, $0<\sigma<1$ is the spread, and $g_{\alpha}/J\in[0.25,0.75]$. All TLS splittings are positive in our numerical calculations. For each chain length, we use the same disorder realizations when varying the mean splitting, the relative spread, or the initial TLS state. For simplicity, we treat $g'_{\alpha}=0$, corresponding to zero detuning in the DQD setting.

	To probe the competition between energy exchange and channel closure, we vary the mean TLS splitting $\bar\Omega$ and compare $T(\bar\Omega)$ with the elastic transmission $T_0$ of the same chain with $g_\alpha=g'_\alpha=0$. Figure~\ref{Fig:T_Omega} shows the geometric mean of $T(\bar\Omega)/T_0$ over disorder realizations at fixed incident electron energy $\mathcal E_{\vex s}=0.5J$. The total energy $E=\mathcal E_{\vex s}+\sum_\alpha s_\alpha\Omega_\alpha$ therefore depends on the initial TLS configuration and its gaps. For every chain length, gap spread, and initial TLS configuration shown, the transmission ratio rises to a maximum at intermediate $\bar\Omega$. In most cases it then dips below unity, indicating stronger reflection than in the same chain with only elastic scattering. At large splittings, $T$ approaches $T_0$.

	In Fig.~\ref{Fig:T_Omega}(a), the size dependence is investigated. The results with different sizes are roughly consistent for $\bar\Omega/J>1$ but show strong size dependence in the small $\bar\Omega/J$ regime. Notably, a larger size leads to a stronger initial growth of transmission, which is consistent with the weak localization intuition: A longer impurity separation tends to suppress reflection more.

	We further analyze the dependence of $\sigma$, concerning the distribution of TLS splittings, in Fig.~\ref{Fig:T_Omega}(b). The three curves nearly coincide for $\bar\Omega/J<1$. At larger $\bar\Omega$, the transmission drops sharply below $T_0$ for $\sigma=0.1$. The minimum is shallower for $\sigma=0.25$, and no clear minimum is visible for $\sigma=0.4$. To understand this trend, consider first the simpler case of equal gaps, $\Omega_\alpha=\bar\Omega$. Exciting one TLS leaves the electron with energy $\mathcal E_{\vex0}-\bar\Omega$. Once this falls below the band edge at $-2J$, every channel with an excited TLS is closed, and a resonance in a closed channel can produce strong elastic reflection, producing a lower transmission. For the finite gap spreads in Fig.~\ref{Fig:T_Omega}(b), the channels with one excited TLS close at different values of $\bar\Omega$. As the spread increases, a resonance associated with one closed channel can occur while other channels still carry transmitted flux. This offers a natural explanation for the weaker transmission minimum at larger $\sigma$.

	In addition, the nonmonotonic transmission seems to be generic for any incident channel. All three initial TLS configurations in Fig.~\ref{Fig:T_Omega}(c), with $n_{\rm exc}=0,5,9$, show a transmission maximum.
	The three curves are essentially the same for $\bar\Omega/J<0.1$. For $n_{\rm exc}=9$, the transmission approaches $T_0$ at smaller mean splittings than for the other two preparations. The large $\bar\Omega$ behavior can be understood by the condition of channel closing, as the incident electron energy $\mathcal E_{\vex s}=0.5J>0$ and the $n_{\rm exc}=9$ channel can only increase the kinetic energy for the inelastic channels, making channel closing much easier than other two cases.

	We emphasize that the quantitative extraction of localization length as a function of $\bar\Omega$ is not established in this work. The main obstacle is the Hilbert space scales exponentially to the number of dynamical impurities, making a proper finite-size extraction not feasible. Nevertheless, our numerical results, the enhancement of total transmission due to dynamical impurities, provide a clear picture of how coherent reflection is impacted by the dynamical impurities.

	\textit{Multi-electron case.---} When multiple electrons are considered, the problem becomes much richer. For example, the TLS impurity can mediate a retarded interaction between two electrons through the transverse coupling, similar to a local oscillator (i.e., the Holstein phonon). For the spinless fermion model here, a dynamical interaction (i.e., redistribution of electron energies) appears at order $g^2$, while a static on-site interaction reduces to a one-body energy shift. For spinful electrons with spin-independent coupling, the induced static on-site interaction at order $g^2$ is attractive for a ground-state TLS and repulsive for an excited TLS.

	At finite electron density, Pauli blocking restricts the final states available to a scattered electron. The occupation of impurity-localized electronic states can also modify the local density and the available scattering processes. The electron density in turn affects the TLS dynamics through Eq.~\eqref{Eq:H_I_TLS}. In a mean-field approximation, the TLS Hamiltonian [Eq.~\eqref{Eq:H_TLS}] acquires the term $g_{\alpha}\bar{n}_{\alpha}\sigma^x_{\alpha}+g'_{\alpha}\bar{n}_{\alpha}\sigma^z_{\alpha}$, where $\bar{n}_{\alpha}=\langle n_{j_{\alpha}}\rangle$ is the average local density. This term can shift the TLS splitting and rotate its eigenbasis, changing the effective couplings. Finally, when energetically allowed, an excited TLS can relax by creating particle--hole excitations in the Fermi sea~\cite{GuineaF1985,KnapM2013}. A detailed study incorporating these many-body effects will be reported elsewhere.

	\textit{Discussion.---} Gate-defined quantum-dot arrays offer a platform for testing our predictions of resonant reflection from a single capacitively coupled DQD and nonmonotonic transmission as the splittings of several DQDs are tuned. Electrical control and readout of a single electron in a DQD have been demonstrated experimentally \cite{PettaJR2004,PeterssonKD2010}. At zero detuning, varying the interdot hybridization $\omega_\alpha$ tunes the splitting $\Omega_\alpha=2|\omega_\alpha|$. To test the resonance in Fig. \ref{Fig:RT_one_close}(a), one could prepare the DQD in its ground state and inject electrons at a fixed energy for which the excitation channel is closed. A plunger gate could then lower the on-site energy of the transport dot adjacent to the DQD, providing the attractive potential needed for resonant reflection. The predicted signature is a dip in transmission whose position shifts as the DQD splitting is varied. With several DQDs coupled to different sites of the chain, sweeping their mean splitting would allow a test of the nonmonotonic transmission in Fig. \ref{Fig:T_Omega}.

	The results here should be generalized to other types of dynamical impurities. For example, a single local harmonic oscillator can produce resonant total reflection~\cite{BrandesT2002}, too. For spatially separated oscillator impurities, we also expect nonmonotonic transmission in comparable parameter regimes. Energy exchange with the oscillators can weaken coherent reflection, while increasing their frequencies eventually suppresses inelastic scattering and recovers the elastic limit at fixed coupling. The ability of each oscillator to absorb several quanta could shift the transmission peak and change its sensitivity to the initial impurity state. A local impurity with only two internal states, which we study here, thus provides a minimal setting for studying how energy exchange and resonant elastic reflection arise from the same coupling.

	\begin{acknowledgments}
		The author is grateful to Sankar Das Sarma and Jay Sau for useful discussion. This work is supported by the Laboratory for Physical Sciences. The author thanks OpenAI for providing research access to ChatGPT and Codex through Academic Researchers Program.

		The author used OpenAI’s ChatGPT and Codex (models 5.6 Sol and 6.0 Astra) to assist with analytical derivations, writing the numerical scattering code, preparing figures, and revising the manuscript. The author specified the model, parameter sweeps, and criteria for selecting disorder realizations. The numerical results were checked for flux conservation, and the transmission curves in Fig. \ref{Fig:T_Omega} were checked against the underlying per-realization data. The author takes full responsibility for the content of this work.
	\end{acknowledgments}

\section*{End Matter}

\nopagebreak[4]

	\appendix

	\section{Lattice retarded Green function}\label{App:G_fcn}

	We are particularly interested in the  noninteracting retarded electron Green function given by
	\begin{align}
		G(j-j';z)\equiv\bra{j}\frac{1}{z-\hat{H}_e}\ket{j'},
	\end{align}
	where $z=\mathcal E+i\eta$, $\eta\rightarrow0^+$, and we have used the lattice translation symmetry. In addition, $\hat{H}_e$ has the inversion symmetry, so $G(j;z)=G(|j|;z)$. For an infinite lattice system,
	\begin{align}
		G(j;z)=\int_{-\pi}^{\pi}\frac{dp}{2\pi}
		\frac{e^{ipj}}{z+2J\cos p}.
		\label{Eq:Green-integral}
	\end{align}
	One can clearly see that only the even in $j$ part survives the integral. To derive the expression of Green function, we introduce $w=e^{ip}$. Thus, $dp =dw/(iw)$ and $2\cos(p)=w+w^{-1}$. Using $n=|j|$, Equation~(\ref{Eq:Green-integral}) becomes
	\begin{align}
		G(j;z)=\frac{1}{2\pi i}\oint\limits_{|w|=1}\frac{w^ndw}{Jw^2+zw+J}=\left[\frac{w^n}{2Jw+z}\right]_{w\rightarrow w_{\rm in}},
	\end{align}
	where $w_{\rm in}$ is the root inside the unit circle. The root is determined by the zeros of $Jw^2+zw+J$:
	\begin{align}
		w_{\pm}=\frac{-z\pm\sqrt{z^2-4J^2}}{2J}.
	\end{align}
	$w_+w_-=1$. One can check that $w_+$ is inside the unit circle. Thus, the Green function is given by
	\begin{align}
		G(j;z)=\frac{\left(\frac{-z+\sqrt{z^2-4J^2}}{2J}\right)^n}{\sqrt{z^2-4J^2}}.
	\end{align}

	To simplify the expression of the retarded Green function, we perform the following parameterization. First, the energy $\mathcal E$ can be expressed by:
	\begin{align}
		\mathcal E=\begin{cases}
			2J\cosh(\kappa),& \text{for }\mathcal E>2J,\\
			-2J\cos(k),& \text{for }-2J<\mathcal E<2J,\\
			-2J\cosh(\kappa),& \text{for }\mathcal E<-2J.
		\end{cases}
	\end{align}
	Concomitantly, we obtain
\begin{multline}
		\sqrt{(\mathcal E+i\eta)^2-4J^2}\\
=\begin{cases}
			2J\sinh(\kappa),& \text{for }\mathcal E>2J,\\
			2iJ\sin(k),& \text{for }-2J<\mathcal E<2J,\\
			-2J\sinh(\kappa),& \text{for }\mathcal E<-2J.
		\end{cases}
	\end{multline}
	The difference in sign for $\mathcal E>2J$ and $\mathcal E<-2J$ is due to the infinitesimal $\eta>0$, which picks up different phases for the two cases.
	We note that one can obtain the $|\mathcal E|>2J$ results from the $|\mathcal E|<2J$ by the following substitutions: $k\rightarrow i\kappa+\pi$ for $\mathcal E>2J$ and $k\rightarrow i\kappa$ for $\mathcal E<-2J$. Finally, the retarded Green function is expressed by
	\begin{align}
		G(j;\mathcal E+i0^+)=\begin{cases}
			\frac{(-1)^{|j|}e^{-\kappa |j|}}{2J\sinh(\kappa)},& \text{for }\mathcal E>2J,\\[2mm]
			\frac{-ie^{ik|j|}}{2J\sin(k)},& \text{for }-2J<\mathcal E<2J,\\[2mm]
			-\frac{e^{-\kappa |j|}}{2J\sinh(\kappa)},& \text{for }\mathcal E<-2J,
		\end{cases}
	\end{align}
	The plane-wave states ($|\mathcal E|<2J$) manifest the $e^{ik|j|}$ factor, while the evanescent states ($|\mathcal E|>2J$) show exponential decays. Interestingly, there is an additional $(-1)^{|j|}$ factor for the $\mathcal E>2J$ case, reflecting the differences between the two branches of evanescent states. We define $F_m\equiv G(0;\mathcal{E}_m+i0^+)$ in the later discussion.

\section{Single-impurity Lippmann--Schwinger equation}\label{App:Single_TLS}

For a single impurity at $j=0$, let $\psi_m(j)$ denote the electron wavefunction in TLS channel $m$, with $\mathcal{E}_m=E-m\Omega$ and $\bar m=1-m$. For a unit-amplitude wave incident from the left in channel $s$, $\mathcal{E}_s=-2J\cos k_s$ with $0<k_s<\pi$. The channel-projected Lippmann--Schwinger equation is
\begin{align}
\psi_m(j)={}&\delta_{ms}e^{ik_sj}+G(j;\mathcal{E}_m+i0^+)\nonumber\\
&\times\left[u_m\psi_m(0)+g\psi_{\bar m}(0)\right].
\label{Eq:LS_channel}
\end{align}
Writing $a_m=\psi_m(0)$ and using $F_m$ defined in Appendix~\ref{App:G_fcn}, the local equations are $a_m=\delta_{ms}+F_m(u_ma_m+ga_{\bar m})$. Solving this two-component system gives
\begin{align}
a_s&=\frac{1-u_{\bar s}F_{\bar s}}{D(E)},\qquad
a_{\bar s}=\frac{gF_{\bar s}}{D(E)},\nonumber\\
D(E)&=(1-u_0F_0)(1-u_1F_1)-g^2F_0F_1.
\label{Eq:LS_local_solution}
\end{align}
For an open outgoing channel, $G(j;\mathcal{E}_m+i0^+)=F_me^{ik_m|j|}$ and $F_m=-i/v_m$. Matching the left- and right-moving waves gives
\begin{align}
t_{s\rightarrow m}&=a_m,\qquad
r_{s\rightarrow m}=a_m-\delta_{ms},\nonumber\\
T_{s\rightarrow m}&=\frac{v_m}{v_s}|t_{s\rightarrow m}|^2,\qquad
R_{s\rightarrow m}=\frac{v_m}{v_s}|r_{s\rightarrow m}|^2.
\label{Eq:LS_flux}
\end{align}
These expressions reproduce the two-open-channel results in the main text. When $\bar s$ is closed, eliminating $a_{\bar s}$ instead gives $t_{s\rightarrow s}=[1+i\mathcal{U}_{\mathrm{eff},s}/v_s]^{-1}$, where $\mathcal{U}_{\mathrm{eff},s}=u_s+g^2/(F_{\bar s}^{-1}-u_{\bar s})$, recovering Eqs.~(\ref{Eq:R_T_one_open}) and (\ref{Eq:U_eff}). At $1-u_{\bar s}F_{\bar s}=0$ and $g\ne0$, the numerator of $a_s$ vanishes while $D=-g^2F_sF_{\bar s}\ne0$. Thus, perfect reflection occurs at the uncoupled-channel bound-state energy, rather than at a bound-state pole of the finite-$g$ problem.

Bound states follow from Eq.~(\ref{Eq:LS_channel}) without the incident-wave term. For $g\ne0$, normalizability requires $|E_b|>2J$ and $|E_b-\Omega|>2J$, and their energies obey
\begin{align}
\left[F_0^{-1}-u_0\right]\left[F_1^{-1}-u_1\right]=g^2,
\label{Eq:LS_bound_state}
\end{align}
with $F_m$ evaluated at total energy $E_b$. This is equivalent to $D(E_b)=0$ with decaying solutions in both channels.

\section{Transfer matrix and numerical calculations}\label{App:Transfer_matrix}

\nopagebreak[4]
At total energy $E$, channel $\vex m$ has electron energy $\mathcal E_{\vex m}=E-\sum_\alpha m_\alpha\Omega_\alpha$. For incident configuration $\vex s$, $E=\mathcal E_{\vex s}+\sum_\alpha s_\alpha\Omega_\alpha$. The calculations in Fig.~\ref{Fig:T_Omega} hold $\mathcal E_{\vex s}=0.5J$ fixed, rather than the total $E$, as the gaps and initial configurations vary.
Let $\boldsymbol\psi_j$ collect the channel amplitudes, $\mathsf E=\operatorname{diag}(\mathcal{E}_{\vex{m}})$, and $\hat{\mathsf{V}}_j$ be the impurity potential matrix specified by Eq.~(\ref{Eq:H_I_TLS}). Writing $\boldsymbol\Phi_j=(\boldsymbol\psi_j^T,\boldsymbol\psi_{j-1}^T)^T$, the lattice equation gives
\begin{align}
\boldsymbol\Phi_{j+1}
=\begin{pmatrix}(\hat{\mathsf{V}}_j-\mathsf E\hat{1})/J&-\hat{1}\\
\hat{1}&0\end{pmatrix}\boldsymbol\Phi_j,
\label{Eq:TM_site}
\end{align}
where $\hat{1}$ is the identity in channel space.

For a clean interval, choose $z_{\vex{m}}+z_{\vex{m}}^{-1}=-\mathcal{E}_{\vex{m}}/J$, with $z_{\vex{m}}=e^{ik_{\vex{m}}}$, $0<k_{\vex{m}}<\pi$, for an open channel and $|z_{\vex{m}}|<1$ for a closed channel. The exact $d$-step clean transfer matrix in each channel is
\begin{align}
\mathsf C_{\vex{m}}^d&=
\begin{pmatrix}
f_{d+1}(z_{\vex{m}})&-f_d(z_{\vex{m}})\\
f_d(z_{\vex{m}})&-f_{d-1}(z_{\vex{m}})
\end{pmatrix},\nonumber\\
f_n(z)&=\frac{z^n-z^{-n}}{z-z^{-1}}.
\label{Eq:TM_clean_power}
\end{align}
At $z=\pm1$, the finite limit is $f_n(z)=nz^{n-1}$. Between consecutive impurity transfers, $d=j_{\alpha+1}-j_\alpha-1$. Thus, clean intervals require no repeated site-matrix multiplication.

Numerically, clean propagation is represented by amplitudes anchored at the interval endpoints, using nonnegative powers of the outgoing or decaying roots. We solve all impurity matching conditions simultaneously as a sparse linear system, keeping every source-connected configuration, including closed channels. The boundary conditions specify one incoming wave from the left and exclude other incoming waves and growing evanescent solutions in the exterior regions. For unit incident amplitude in channel $\vex{s}$, the outgoing amplitudes $t_{\vex{m}}$ and $r_{\vex{m}}$ give
\begin{align}
T&=\sum_{\vex{m}\in\mathrm{open}}\frac{v_{\vex{m}}}{v_{\vex{s}}}|t_{\vex{m}}|^2,\nonumber\\
R&=\sum_{\vex{m}\in\mathrm{open}}\frac{v_{\vex{m}}}{v_{\vex{s}}}|r_{\vex{m}}|^2,
\label{Eq:TM_probabilities}
\end{align}
where $v_{\vex{m}}=\sqrt{4J^2-\mathcal{E}_{\vex{m}}^2}$. Note that the closed channels carry no asymptotic flux. We have checked the flux conservation, $T+R=1$, for all the numerical outputs.

For each realization in Fig. 3, we chose nine distinct impurity sites uniformly at random. Keeping these sites fixed, we sampled the on-site potentials $U_\alpha$ independently and uniformly from $[-2J,2J]$ and retained the first set satisfying the transmission cutoff specified in the main text. We sampled $g_\alpha/J$ uniformly from $[0.25,0.75]$ and obtained each gap profile $z_\alpha$ by centering nine independent uniform random values from $[-1,1]$ about their sample mean and rescaling them to unit root mean square.

\end{document}